\documentclass[9pt]{article}
\usepackage{spconf,amsmath,graphicx,mathtools,bm}
\usepackage{tikz}
\usepackage{amsfonts}
\usepackage{multicol}
\usepackage{graphics}
\usepackage{caption}

\title{Bifocal Neural ASR: Exploiting Keyword Spotting for Inference Optimization}
\name{Jonathan Macoskey, Grant P. Strimel, Ariya Rastrow}
\address{Amazon.com\\{\small \tt$\{$macoskey,gsstrime,arastrow$\}$@amazon.com }}
\begin{document}
	%
	\maketitle
	\small
	\begin{abstract}
		We present Bifocal RNN-T, a new variant of the Recurrent Neural Network Transducer (RNN-T) architecture designed for improved inference time latency on speech recognition tasks.
		The architecture enables a dynamic pivot for its runtime compute pathway, namely taking advantage of keyword spotting
		to select which component of the network to execute for a given audio frame.
		To accomplish this, we leverage a recurrent cell  we call the Bifocal LSTM (BF-LSTM), which we detail in the paper.
		The architecture is compatible with other optimization strategies such as quantization, sparsification, and applying time-reduction layers, making it 
		especially applicable for deployed, real-time speech recognition settings.
		We present the architecture and report comparative experimental results on voice-assistant speech recognition tasks.
		Specifically, we show our proposed Bifocal RNN-T can improve inference cost by 29.1\% with matching word error rates and only a minor increase in memory size.
	\end{abstract}

	\section{Introduction}
	\label{sec:intro}
	The increasing omnipresence of smartphones, smart speakers, and 
	tablets coupled with the adoption of voice assistants has motivated a modern trend
	to develop Automatic Speech Recognition (ASR) systems which fully operate on local devices \cite{McGraw2016, JinyuLiRuiZhaoHuHu2019, Sainath2020}.
	The promise of on-device ASR includes increased reliability, improved latency and privacy benefits by alleviating the need to stream audio to servers.
	In order to realize these benefits, however, new approaches are required to address the challenges 
	posed by compute constrained devices.
	
	A complementary trend prevalent in the literature is a shift from
	traditional Hidden Markov Model
	(HMM)-based ASR systems to end-to-end approaches \cite{Graves2012, Graves2013, Graves2014, Chan2016, Soltau2017}.
	These end-to-end architectures replace the typically disjoint components in an ASR system with a single, fully neural architecture trained over large amounts of data.
	
	The fully neural approaches are strong candidates for low-footprint settings due to their simplicity
	and uniform compression ability; however, when deployed on devices with hardware constraints, e.g. limited compute and memory bandwidth, they still require careful use of compression algorithms and optimization techniques to achieve real-time, low-latency speech recognition.

	In order to address this latency bottleneck, several engineering and modeling techniques have 
	been proposed, especially for streaming architectures such as RNN-T \cite{Graves2012}.
	Several studies \cite{Alvarez2016, Mishchenko2018,Nguyen2020} have investigated moving from 32-bit weights down to
	8 or 4 bits for neural ASR models.
	These quantized networks not only reduce model size but also
	eliminate costly floating point operations and can better integrate with hardware \cite{Yang2018}.
	Likewise, experiments have shown that low-rank matrix factorization \cite{Zhu2018} 
	and network sparsification methods \cite{Zhu2018, Pang2018} can be applied to an extent without significant accuracy degradation.
	Reducing parameter operations and complexity leads to inference cost gains for these methods.
	
	Higher-level architectural changes 
	designed to improve inference speed have also been proposed \cite{Shangguan2019}. 
	The RNN topologies of the CIFG-LSTM \cite{Greff2017}, which has 25\% fewer parameters compared
	to the standard LSTM \cite{Hochreiter1997}, and Simple Recurrent Units (SRU) \cite{Lei2020} are suggested as 
	alternative recurrent layers for on-device ASR.
	These authors show a retention in modeling capacity 
	with fewer parameters and less compute required at runtime.
	Finally, there are common latency reduction methods, such as time-reduction \cite{Chan2016,Soltau2017, He2018},
	which reduce the effective decoding frame rate at which audio is processed.

	In this work, we present a new variant of the RNN-T architecture, Bifocal RNN-T, which is designed to lower latency for on-device ASR applications.
	The architecture leverages design characteristics common to voice assistant applications, specifically keyword spotting. 
	Voice assistants will typically use a lightweight, continuously-running keyword spotter model strictly to detect a pre-determined wake word (WW) or phrase (e.g. ``\emph{Alexa}" or ``\emph{Ok Google}") \cite{Sun2017,Arik2017}.
	For efficiency and privacy, ASR decoding is deferred until WW recognition, at which point ASR begins decoding streamed audio along 
	with a buffered lead-in segment of pre-WW audio.
	While this design provides a natural user experience with additional robustness built in (i.e. lowering the false-accept rate by re-verifying the presence of the WW during ASR), the buffering introduces a latency hurdle which generic techniques do not address.
	Our Bifocal RNN-T architecture is constructed to address this problem by incorporating keyword spotting as part of its design.
	Our contributions are complementary to the existing literature and can be used in conjunction with the group of optimizations referenced.
	The remainder of the paper is organized as follows. Sections \ref{sec:bifocal_rnnt} and \ref{sec:bifocal_lstm}  detail the Bifocal RNN-T and its critical components, namely the Bifocal LSTM. 
	Section \ref{sec:interleaving} extends the basic approach further while Section \ref{sec:experimental_results} outlines experiments that show the benefits of Bifocal RNN-T on speech recognition tasks.

	\vspace{-1.0mm}
	\section{Bifocal RNN-T}
	\label{sec:bifocal_rnnt}
	The Recurrent Neural Network Transducer (RNN-T) \cite{Graves2012} is a fully neural, sequence-to-sequence architecture that is particularly applicable for ASR modeling.
	The architecture is trained end-to-end with what can be viewed as three logical components:
	the encoder (also referred to as the transcription network),  
	the decoder (also referred to as the prediction network),
	and an optional joint network.
	For ASR, the transcription network $\mathcal{F}$ is a multi-layer RNN, which operates over feature vectors $x_{1:T}$
	extracted from a raw audio signal. For speech, these will be features given by
	transformations over the signal (e.g. log-mel filter bank) and stacking consecutive samples into frames.
	The encoder outputs a vector $h_t^{\text{enc}}$ for each frame of the input based on the sequence of frames observed up to that point:
	\[
	h_t^{\text{enc}} = \mathcal{F} \left( x_{1:t} \right).
	\]
	The decoder $\mathcal{G}$ is also a multi-layer RNN which operates over the sequence of outputted labels $y_{1:M}$.
	The labels in our setting are word pieces. 
	The decoder computes a new vector $h_m^{\text{dec}}$ for each new outputted label of the RNN-T:
	\[
	h_{m}^{\text{dec}} = \mathcal{G} \left( y_{1:m} \right).
	\]
	The prediction network embodies the role of the language model in a traditional HMM hybrid system, operating over labels to compute scores to be
	combined with the output from $\mathcal{F}$.
	This operation is the role of the joint network $\mathcal{J}$, which serves either as (i) a simple additive operation between the outputted vectors of each network \cite{Graves2012} or (ii) as a feedforward network \cite{JinyuLiRuiZhaoHuHu2019}, to construct a probability distribution for the next output label:
	\begin{align*}
		\mathcal{J} \left( h_t^{\text{enc}} , h_{m}^{\text{dec}} \right) = 
		\begin{cases}
			h_t^{\text{enc}} +  u_m^{\text{dec}} & \text{(i)}\\
			\psi \left( W h_t^{\text{enc}} +  V  h_m^{\text{dec}}\right) & \text{(ii)}
		\end{cases} 
	\end{align*}
	where $\psi$ is an activation. 
	Finally, the conditional output distribution $P\left(\hat{y}_{m+1}|x_{1:t}, y_{1:m}\right)$ is obtained by applying a softmax to the result of $\mathcal{J}$.
	For real-time speech settings, the $\mathcal{F}$ and $\mathcal{G}$ are commonly built by stacking multiple unidirectional LSTMs \cite{JinyuLiRuiZhaoHuHu2019}
	and the network is trained with what amounts to an extension of the Connectionist Temporal Classification (CTC) approach \cite{Graves2006},
	allowing RNN-T to implicitly learn to align audio and transcriptions.
	
	\begin{figure}[t]
	   \captionsetup{font=small}
		\begin{center}
			\input{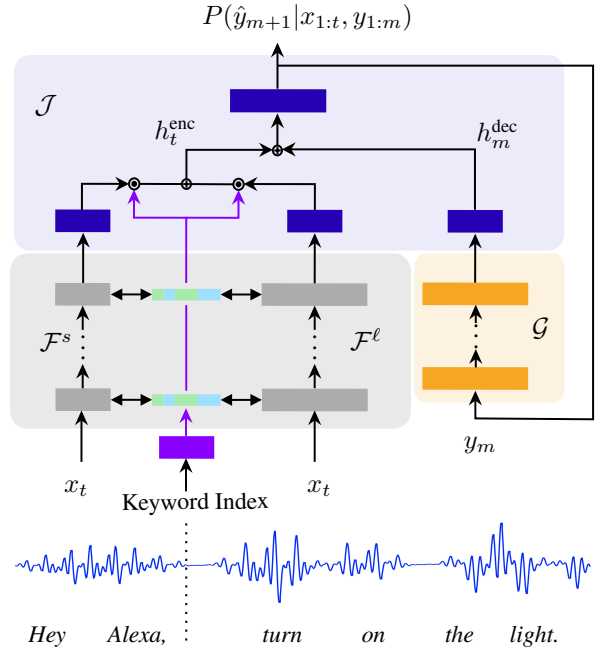}
			
			\vspace{-2.0mm}
			\caption{Bifocal RNN-T architecture. Two encoder networks (a small $\mathcal{F}^s$ and large $\mathcal{F}^{\ell}$) are trained using BF-LSTM cells (gray), see Section \ref{sec:bifocal_lstm}.
				Their mutually exclusive execution is determined by the frame index signaling the end of the WW (purple). The hidden state is transitioned between encoders after the WW. The decoder (gold) and the joint network (blue) operate as in standard RNN-T.
			}
			\vspace{-7.0mm}
			\label{fig:bifocal_rnn_t}
		\end{center}
	\end{figure}
	
	\begin{figure*}[th!]
	\captionsetup{font=small}
		\centering
		
		\input{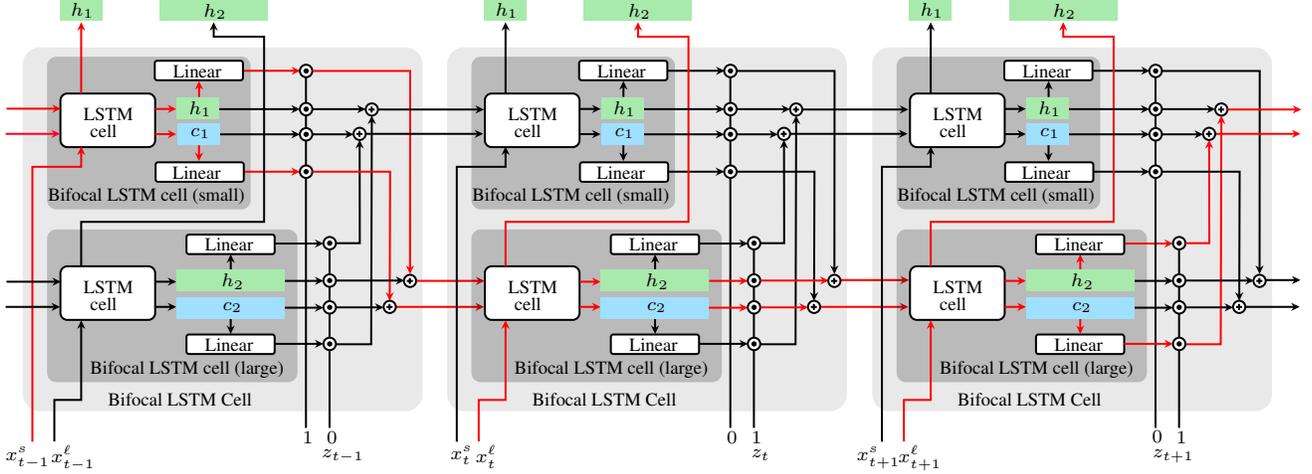}
		
		\vspace{-2.0mm}
		\caption{Unfolded Bifocal LSTM through time. Each block represents the execution of the Bifocal LSTM Cell at single time step. While the layer is fully connected and differentiable, only one component LSTM cell is executed at each timestep during inference. This is highlighted by the red pathways and is controlled by the switching signal $z_t$. 
		}
		\vspace{-6.0mm}
		\label{fig:bifocal_lstm_cell}
	\end{figure*}

	Voice assistant enabled devices, like Amazon Echo and Google Home, use keyword spotting as a central design element of their user interface and entry point to ASR model decoding.
	A small, highly accurate keyword spotting model is used to continuously process audio on-device and, when appropriate, activate streaming audio
	through the ASR module.
	A voice assistant system, however, can buffer a segment of audio to allow any ``lead-in'' words spoken before the WW (and including WW itself) to also be decoded by the ASR module.
	For example, in utterances ``\emph{Hey Alexa, turn on the light.}'' or ``\emph{Hello Echo, is it raining?}'', the lead-in segment would be
	``\emph{Hey Alexa}'' or ``\emph{Hello Echo}'', respectively.
	
	Here we propose a modified RNN-T architecture which takes advantage of the keyword signal for low compute settings.
	We call our approach Bifocal RNN-T because the architecture has multiple ``lenses'' designed to focus on different segments of an utterance.
	The distinguishing feature of the design is training two alternative encoder RNNs and leveraging the keyword spotting to pivot between the two.
	A small $\mathcal{F}^s$ encoder network is trained for the lead-in segment and a large $\mathcal{F}^\ell$ encoder network 
	is trained for processing the remainder of the audio.
	While sharing an equal number of layers, $\mathcal{F}^s$ has a smaller number of hidden units than $\mathcal{F}^\ell$, allowing a faster but coarser processing of frames.
	$\mathcal{F}^\ell$, in contrast, will have a higher capacity but at the cost of a higher compute. 
	The natural lower perplexity
	of the spoken words prior to the WW lends itself to using a lighter-weight model while the higher perplexity of the post-WW utterance lends itself to using a model with greater capacity.
	The result of employing this dual-encoder Bifocal architecture is that we obtain a larger model with a lowered total compute cost summed over all frames of an utterance. 
    Note that the Bifocal design places an emphasis on reducing the encoder inference cost since it 
	is a primary bottleneck during runtime; the encoder is larger, executed every frame, and not cacheable unlike the decoder which is smaller, requires fewer queries (which also can be parallelized across different threads), and whose results are cacheable during decoding.	
	
	Because the lead-in segment is buffered, it becomes critical to process this audio backlog rapidly in order for the processing to catch-up and finish decoding in real-time.
	Failure to do so risks perpetual lag during inference, which results in a significant response latency.
	Moreover, the lead-in segment will typically amount up to 30\% of the total audio.
	
	Figure \ref{fig:bifocal_rnn_t} depicts the Bifocal RNN-T with its 
	two encoder branches whose execution is toggled by the frame index denoting the completion of the WW, where the index is passed to the ASR module from the upstream keyword spotter component. 
	We build the $\mathcal{F}^s$ and $\mathcal{F}^\ell$ encoders with
	a specialized LSTM cell, detailed in the next section, which enables us to train both networks in unison
	and provide a mechanism to smoothly transition the hidden state of the RNN-T network during the switch after the WW. \vspace{-2mm}
	
	\section{A Switching LSTM Cell}
	\label{sec:bifocal_lstm}
	
	The critical component of the Bifocal RNN-T is the Bifocal LSTM (BF-LSTM).
	The BF-LSTM is a trainable RNN cell that enables dynamic inference pathways at runtime.
	The design is implemented by training two distinct LSTM cells per layer that learn to transfer their hidden 
	states into one another by projection.
	We refer to these two cells working in tandem within the BF-LSTM as the \textit{small} 
	and \textit{large} cells.
	Their names reference their relative dimensionality.
	The small cell has fewer units, which amounts to a faster inference with a less granular modeling function.
	The large cell possesses a greater capacity but at the cost of more parameters on which to operate.
	These cells have different hidden state sizes, so to switch between them at different frames, we
	add learned ``translation'' operations for projecting into each others' state spaces.
	Switching at each timestep is controlled by a user-provided input $z_t \in \left\{0,1\right\}$ to signal the small ($z_t =0$) and large ($z_t = 1)$ cells. 
	We present the equations for the BF-LSTM below (note the affine transforms are given without their bias for brevity but can be included).\vspace{0.5mm}

	\def\lstm[#1][#2][#3]{\overbrace{f_t^{#1} = \sigma ( W_f^{#1} x_t^{#1} + U_f^{#1} h_{t-1}^{#1} )}^\text{Standard LSTM (#2) } \\[#3]
		i_t^{#1} = \sigma ( W_i^{#1} x_t^{#1} + U_i^{#1} h_{t-1}^{#1}  ) \\[#3]
		o_t^{#1} = \sigma ( W_o^{#1} x_t^{#1} + U_o^{#1} h_{t-1}^{#1} ) \\[#3]
		\tilde{c}_t^{#1} = \tanh ( W_c^{#1} x_t^{#1} + U_c^{#1} h_{t-1}^{#1} ) \\[#3]
		c_t^{#1} = f_t^{#1} \circ c_{t-1}^{#1}  + i_t^{#1} \circ \tilde{c}_t^{#1} \\[#3]
		h_t^{#1} = o_t^{#1} \circ \sigma  ( c_t^{#1} )}
	
	\noindent
	\resizebox{8.4cm}{!}{
		\noindent
		\begin{minipage}{\linewidth}
			\begin{multicols}{2}
				\noindent
				\begin{align*}
					\lstm[s][$s$mall][2.5pt]
				\end{align*}
				\begin{align*}
					\lstm[\ell][$\ell$arge][0pt]
				\end{align*}
			\end{multicols}
			\vspace{-7.5mm}
			\begin{align*}
				\overbrace{\hat{c}_t^\ell = P_c^s c_t^s \qquad   \hat{c}_t^s = P_c^\ell c_t^\ell }^\text{State Projections}\\
				\hat{h}_t^\ell = P_h^s h_t^s  \qquad  \hat{h}_t^s = P_h^\ell h_t^\ell \\
			\end{align*}
			\vspace{-11.5mm}
			\begin{align*}
				c_t^s := c_t^s  \left(1 - z_t \right) + \hat{c}_t^s z_t & \qquad c_t^\ell := \hat{c}_t^\ell z_t + c_t^\ell \left(1 - z_t \right)\\
				h_t^s := h_t^s \left(1 - z_t \right) + \hat{h}_t^s  z_t & \qquad h_t^\ell :=  \hat{h}_t^\ell z_t + h_t \left(1 - z_t \right)\\
			\end{align*}
			\vspace{-2.5mm}
		\end{minipage}
			\vspace{-2.5mm}
	}\vspace{-5.5mm}
	The learned state projection matrices unique to the BF-LSTM cell, $P_c^s, P_h^s \in  \mathbb{R}^{h^s \times h^\ell}$ and $P_c^\ell, P_h^\ell \in  \mathbb{R}^{h^\ell \times h^s}$, produce the translated states $\hat{c}_t^\ell, \hat{h}_t^\ell \in \mathbb{R}^{h^\ell}$ and $\hat{c}_t^s, \hat{h}_t^s \in \mathbb{R}^{h^s}$, respectively,
	which are toggled and then combined to rewrite the state vectors based on the switching variable $z_t$ designating the path of inference execution at timestep $t$.
	
	The above construction is a fully differentiable recurrent layer which allows for straightforward training with standard backpropagation through
	time methods with all pathways computed.
	However, during inference, only one of the sub-cells' computation needs to be executed on each frame based on $z$.
	Furthermore, the state projection
	operations only need to be carried out during a switch where $z_t \neq z_{t+1}$. Figure \ref{fig:bifocal_lstm_cell} shows the fully connected unit through time and highlights a single pathway for inference.
	
	One observes that there are no restrictions on the input dimensions (for $x^s$ and $x^\ell$) and output dimensions (for $h^s$ and $h^\ell$);
	all sizes can differ.
	We use this inherent flexibility to build our Bifocal RNN-T encoders by simply stacking several BF-LSTM layers in sequence.
	While the first layer has matching input dimensions of the audio features, the output dimensions (and thus input dimension of subsequent layers) will have non-matching dimensions.
	A final projection layer is used to map the two outputs from the encoder branches onto a matching dimensionality. 
	The projected outputs are likewise combined using the $z_t$ switch to forward a single output to be used for the joint network. 
	
	\section{Interleaving}
	\label{sec:interleaving}
	While we have thus far presented a design for switching based on pre/post-WW,
	the BF-LSTM, as seen in Section \ref{sec:bifocal_lstm}, is able to train over arbitrary switching patterns that translate in both directions.
	Moreover, the BF-LSTM design can be generalized beyond just two pathways. 
	At the cost of additional projection parameters, the method can support an arbitrary number of branches.
	We therefore extended our Bifocal RNN-T approach to a regime we term \textit{interleaving}.
	
	Interleaving still relies on a small encoder network for lead-in processing,
	but we also attempt to train a set of encoders $\mathcal{F}^{\ell_1}, \mathcal{F}^{\ell_2}, \dots, \mathcal{F}^{\ell_k}$ which are interchangeable for 
	post-WW processing.
	While additionally increasing the overall size of the model, 
	the dimensionality of these encoders and their execution
	schedule can be chosen to lower the compute requirements across the full sequence frames.
	In a ``Trifocal'' setting, which we adopt for our experiments,
	a large encoder $\mathcal{F}^{\ell_l}$ and small encoder $\mathcal{F}^{\ell_s}$ share the post-WW processing burden
	by switching back-and-forth according to a predetermined pattern.
	For example, the schedule $(\ell_l, \ell_l, \ell_s, \ell_s)$ would cycle through each consecutive sequence 
	of four frames by using $\mathcal{F}^{\ell_l}$ for two frames, then projecting to use $\mathcal{F}^{\ell_s}$ for two frames 
	before projecting back to $\mathcal{F}^{\ell_l}$'s state space to repeat the pattern.
	In Section \ref{sec:experimental_results}, we experiment with several of these schedules and report their improved inference cost but adverse impact on the model's predictive performance.
	We note here that interleaving shares a similar approach to those presented in \cite{Graves2016, Jernite2017} which
	leverage variable compute for RNN applications such as text-based character prediction and music modeling. \vspace{-2.5mm}
	
	\section{Experimental Results}
	\label{sec:experimental_results}
	We investigate the model performance for the Bifocal RNN-T architecture on a production voice assistant ASR task.
	The models are trained using teacher forcing with 42k hours of audio consisting of de-identified utterances of far-field, English-locale, virtual assistant tasks.
	Tasks span all Alexa domains including contacts, home automation, music, etc.
	The data consists only of utterances with the WW present, labeled with a WW frame index generated by a pre-existing keyword spotting model.
	An average of 31.8\% of each utterance consists of lead-in segment data. 
	A typical utterance in this dataset consists of 260 frames of audio sampled at 16 kHz. 
	Acoustic features are extracted using log-Filterbank Energies (LBFE) with 64 dimensions.
	Feature frames are downsampled by a factor of 3 and are stacked with a 
	stride size of 2 to produce an overall frame size of 30ms. 
	
	Our baseline RNN-T model is built with 
	five LSTM encoder layers with 1024 units per layer, two LSTM decoder layers with 1024 units per layer, and an additive joint network with no trainable parameters
	resulting in an RNN-T model with 63.5M total parameters, 42.7M of which belong to the encoder.
	Word piece tokens were generated by extracting a vocabulary of the 4k most frequent subword units (plus a blank symbol) using
	a unigram language model \cite{Kudo2018}.
	We also build a smaller baseline model with an encoder using $852$ units per encoder layer which will match the computational cost 
	of our Bifocal model.
	
	We compare against a Bifocal RNN-T, which includes the above specifications of our baseline RNN-T (using BF-LSTM in place of LSTM encoder layers) plus a second encoder network, 
	the lead-in encoder, $\mathcal{F}^{s}$, consisting of five 256-unit BF-LSTM layers.
	As described in Section \ref{sec:bifocal_rnnt}, the lead-in encoder is only used during the lead-in portion of an utterance. After the lead-in segment of an utterance, the encoder states are projected
	into the dimension of the larger, 1024-unit encoder, $\mathcal{F}^{\ell}$, which evaluates the remainder of an utterance. In this configuration, the projection layers
	are used only once during an utterance.
	We also include results for a Bifocal RNN-T without the use of state projections in order to measure their impact on predictive performance against basic zero initialization. Last, we train Trifocal RNN-T models with different interleaving patterns. Each Trifocal model consists of
	all components of the above Bifocal RNN-T and also includes a third encoder network, $\mathcal{F}^{\ell_s}$, consisting of five 256-unit BF-LSTM layers.
	For all Trifocal models, the lead-in encoder is used during the lead-in segment of an utterance, after which
	the states are projected
	into the large encoder, which evaluates the next frame.
	For the remainder of an utterance, frames are evaluated by either the larger $\mathcal{F}^{\ell_l}$ or smaller $\mathcal{F}^{\ell_s}$ post-WW encoders according to a pre-defined schedule.
	Three Trifocal schedules we tested are shown in Table \ref{tab:schedules}.
	
	\begin{table}[h!]
	    \captionsetup{font=small}
		\small
		\centering
		\begin{tabular}{ll}
			\hline
			Model      & Schedule                                           \\ \hline
			Trifocal A & $(\ell_l, \ell_l, \ell_s, \ell_s)$                         \\
			Trifocal B & $(\ell_l, \ell_s, \ell_s)$                \\
			Trifocal C & $(\ell_l, \ell_l, \ell_s, \ell_s, \ell_s, \ell_s)$ \\ \hline
		\end{tabular}
		\caption{Trifocal model schedules.}
		\label{tab:schedules}
		\vspace{-5mm}
	\end{table}
	
	Each model is evaluated on a hold out test set of
	utterances for virtual assistant tasks. Our predictive metric is word error rate (WER), and we decode using a standard beam search with a beam size of 16.
	To measure compute cost, we calculate the total number of encoder floating point operations (FLOPs) required to analyze each utterance.
	This calculation includes the total number of operations processed by each encoder in addition to the operations associated with the projection of state variables where required. 
	Again, our focus on the encoder is because it presents the critical latency bottleneck for real-time decoding.
	The performance and compute cost for each model are shown in Table \ref{tab:results}.
	\begin{table}[h!]
	    \captionsetup{font=small}
		\small
		\resizebox{\columnwidth}{!}{%
		\centering
		\begin{tabular}{lrrrr}
			\hline
			Model      & WER  & Params & FLOPs & Cost Reduct. \\ \hline
			Baseline   & -     & 42.7M  & 11.1B & -              \\
			Baseline  Small  & +3.4\%     & 30.2M  & 7.86B & 29.1\%  \\
			\textbf{Bifocal}    & \textbf{-1.9\%} & \textbf{48.9M}  & \textbf{7.86B} & \bf{29.1\%}\\
			Bifocal (No Proj.)   & +3.4\% & 46.3M  & 7.85B & 29.2\% \vspace{1mm}  \\ 
			Trifocal A & +23.6\%     & 55.1M  & 4.74B & 57.3\%         \\
			Trifocal B & +27.0\%     & 55.1M  & 3.70B & 66.7\%         \\
			Trifocal C & +28.1\%     & 55.1M  & 3.47B & 68.7\%         \\ \hline
		\end{tabular}
		}
		\caption{Model WER and RNN-T encoder compute cost. WER is recorded in relative terms against a standard baseline model. }
		\label{tab:results}
	    \vspace{-3mm}
	\end{table}\\
	The key takeaway from our experiments is that even though the Bifocal encoder is $\sim$10\% larger, it requires $\sim$30\% fewer computations than the baseline.
	Meanwhile, the Bifocal model outperforms the baseline, small baseline, and No Projection Bifocal RNN-Ts, besting the small RNN-T's WER by 5\% while matching its FLOPs. 
	
	All three Trifocal models continued to improve upon compute cost but clearly pay a strikingly steep price in accuracy degradation.
	Despite this, we find it essential to include the Trifocal degradation results because they highlight two important considerations.
	First, they emphasize that the central design feature of the Bifocal RNN-T, utilizing keyword spotting, 
	is justified as an effective way to switch between encoders. Arbitrary switching mid-stream, like the Trifocal models, is unlikely to retain accuracy.
	Second, the difference in WER between Trifocal A and that of B and C
	shows it is not the frequency of encoder switching but rather when and what fraction of frames are processed using the small encoder (A's 50\% post-WW compared to B and C's 66\%) that is the determining factor in predictive performance. This observation reinforces that success is dictated by where to strategically use the small encoder like we do with the Bifocal design. \vspace{-2.5mm}

	
	\section{Conclusion} \vspace{-2.5mm}
	\label{sec:conclusion}
	We present an extension of the RNN-T architecture, Bifocal RNN-T, designed for improved inference-time latency for on-device ASR.
	By exploiting keyword spotting, we show that the Bifocal architecture improves encoder inference cost by 30\% while matching baseline predictive performance.
	The technique is flexible and can be combined with other latency reduction techniques, such as sparsification and quantization.
	In future work we would like to see the approach extended to the decoder, experimented with different topologies, and tried with
	alternative and learnable switching schedules. 

	\let\oldbibliography\thebibliography
	\renewcommand{\thebibliography}[1]{%
		\oldbibliography{#1}%
		\setlength{\itemsep}{0pt}%
	}

	\bibliographystyle{IEEEbib}
	\bibliography{strings,refs}
	
\end{document}